# Work on PETS Developed at CIEMAT

L. Sánchez[1], F. Aragón[1], J. Calero[1], D. Carrillo[1], D. Gavela[1], J.L. Gutiérrez[1], A. Lara[1], E. Rodríguez[1] F. Toral[1],
S. Doebert[2], G. Riddone[2] and A. Samoshkin[2].

1 – Centro de Investigaciones Energéticas, Medioambientales y Tecnológicas
(CIEMAT) – Accelerator Technology Group.
Av. Complutense 40, Madrid – Spain.

2 – European Organization for Nuclear Research (CERN),
CH-1211, Geneva 23, Switzerland.

CIEMAT has been working on the RF power extractor so-called PETS (Power Extraction and Transfer Structure) for the CLIC Test Facility 3 (CTF3) since 2007. The first contribution has been installed at the Test Beam Line (TBL). Additionally, a new PETS configuration is presently under fabrication at CIEMAT and will be installed in the Test Module at CTF3. This paper describes the PETS prototypes design, fabrication and assembly techniques. The characterization of the devices with low RF power is also described.

## 1 Introduction.

PETS are structures to transfer power developed for the future linear collider CLIC. Their main function is to extract power from the drive beam and transfer it to the main beam [1]. When the particles pass through the PETS, interact with the corrugated structure converting the beam kinetic energy into electromagnetic energy. The RF generated power flows through the structure and is collected at the end of the structure by a power extractor and carried through waveguides.

With the aim to demonstrate the feasibility of the CLIC concept, the experimental facility CTF3 is being implemented at CERN. The most important goal of the TBL experiment is to study and validate the drive beam stability during deceleration. The Line will have 16 PETS tanks installed in its final stage [2, 3]. Each TBL PETS will produce the nominal CLIC power of 135 MW with a beam current of 28 A. TBL PETS has a length 4 times longer than CLIC PETS (to compensate the lower drive beam current in CTF3) although the nominal frequency is the same, 12 GHz. CIEMAT is contributing to TBL in the development of a PETS prototype, the assembly and low power testing of 3 PETS more and, finally, the fabrication of several parts for 12 structures more.

Additionally to TBL, the development of CLEX modules in the CTF3 framework was motivated by the need to validate the two-beam acceleration design in a relativistic environment [4]. The Double Length CLIC PETS that CIEMAT has developed and is currently under fabrication corresponds to the prototype designed to generate RF power in the CLIC Experimental Area (CLEX). It is based on a compact concept developed previously by CERN and tested at SLAC.

## 2 CIEMAT contribution to TBL.

The RF design of the TBL PETS was developed at CERN [5]. Based on this design, CIEMAT carried out the engineering design, fabrication and low power RF tests of the first prototype [6]. Additionally, regarding the series production, CIEMAT has contributed by means of the implementation of several modifications in the design, assembly and low power testing of three structures and also with the fabrication and brazing of several parts such as cooling pipes, waveguides and flanges.

### 2.1 PETS TBL first prototype.

*2.1.1 Engineering Design.*

Figure 1 shows a general scheme of the TBL PETS first prototype. The PETS consists of eight copper rods clamped by screws and aligned by dowel pins. As the sag of the rods due to its own weight is estimated as 19 µm in the case of a rod in horizontal position, a steel ring is placed surrounding the bars for rod sagging compensation. At the end of the bars, a power extractor is placed. It consists of three brazed parts and two rectangular WR90 waveguides. The waveguides misalignment with respect to the vacuum tank wall is absorbed by bellows.

The copper rods are conduction cooled by four copper cooling pipes. Custom vacuum tight connectors with Helicoflex seals were designed, and joined to the cooling pipes by vacuum brazing.

The structure is supported by a vacuum tank of stainless steel, in which the ports and bellows are TIG welded and the caps are Conflat flanges.

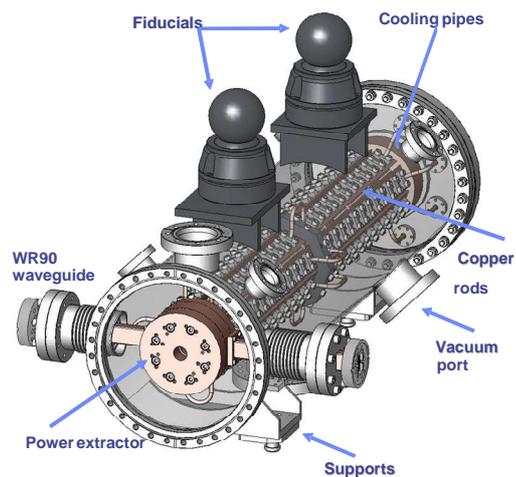

Figure 1. Artistic view of TBL PETS structure

The position of axis structure is mechanically transferred to the external fiducials

*2.1.2 Fabrication and Assembly.*

a) Copper rods. One of the most challenging issues was the fabrication of the copper rods, due to the tight shape tolerance, ± 20 µm, and roughness, less than 0.4 µm. They were made of OFE Cu and machined by high speed milling. Two intermediate stress relieve heat treatments at 180ºC

LCWS11

were carried out. Figure 2 shows one of these copper rods.

The shape of the rods was checked with a 3-D measurement machine, obtaining that most of cells were within tolerances. The vertical deviation due to the sagging of the rods was compensated by means of the intermediate steel ring.

b) Coupler. The coupler has a choke at the beam hole to reflect the power generated by the corrugated structure and two waveguides for extracting that. It is composed of three OFE Cu parts joined by vacuum brazing at CERN. Table I summarizes the brazing parameters.

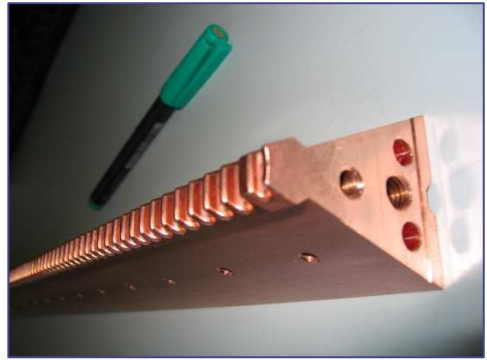

Figure 2. Copper rod of the TBL PETS first prototype

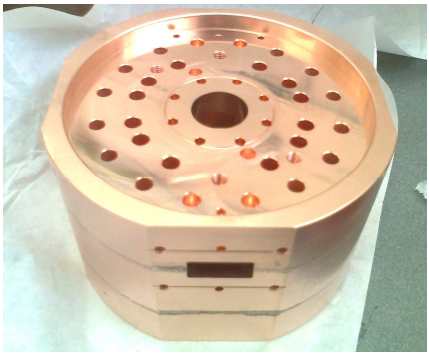

Figure 3. Brazed coupler of the TBL PETS first prototype.

Table I: Parameters for the brazing of the coupler

| Filler alloy (%wt) | Filler melting range (ºC) | Brazing temp (ºC) | Vacuum (mbar) |
|---|---|---|---|
| 68.5 Ag/26.5 Cu/5Pd | 807-810 | 823 | <10-5 |

The alignment of the disks of the power extractor is specially demanding, since the required concentricity is +/- 10 μm, and the filler can not overflow into RF volumes, especially in the choke area, as it would change the power reflection efficiency. Figure 3 shows the brazed parts.

c) Waveguides and cooling pipes. Two rectangular waveguides extract the power from the coupler. The waveguide assembly is composed of the waveguide pipe, an OFE Cu flange, a high vacuum CF flange and a RF flange, both of stainless steel 316 LN. The group was joined by vacuum brazing using an Au-Ni alloy as filler.

The cooling pipes were also joined by brazing. In this case, stainless steel connectors were Ni plated with the purpose of increasing filler wettability on the stainless steel, since the chromia scale which protects against corrosion is even stable at operation conditions [7]. These parts were brazed using the parameters summarized in Table II. After brazing, a helium leak test was carried out on the cooling circuit and waveguide/flanges parts to verify the vacuum tightness.

Table II. Summary of the brazing parameters for the TBL PETS brazed parts

|  | Filler alloy (%wt) | Filler melting point (ºC) | Brazing temp (ºC) | Vacuum (mbar) | Surface treatment |
|---|---|---|---|---|---|
| Waveguides | 82Au/18 Ni | 950 | 980 | <10-5 |  |
| Cooling pipes | 72Ag/28Cu | 780 | 800 | <10-5 | Ni plating on SS |

A special tooling was developed to bend the cooling pipes in two stages: the first bend to half the length of the final geometry, to fit in the size of the brazing furnace, and the latter bend, to the final shape.

d) Assembly. The assembly was in vertical position and started positioning the copper rods at the end support using dowel pins and screws (Figure 4). Once all the copper rods were positioned, the distance between the flat sides of opposite rods was checked and the power extractor was placed at the top of the bars (Figure 5). Afterwards, the cooling pipes were fixed with clamps to guarantee the good thermal contact necessary for the conduction cooling. The first prototype was equipped with temperature sensors (PT-100 type) compatible with UHV conditions. Finally, the tank wall was assembled and the waveguides fixed. Once the tank was closed, helium leak test was performed showing a leak rate below $10^{-10}$ mbar·l/s

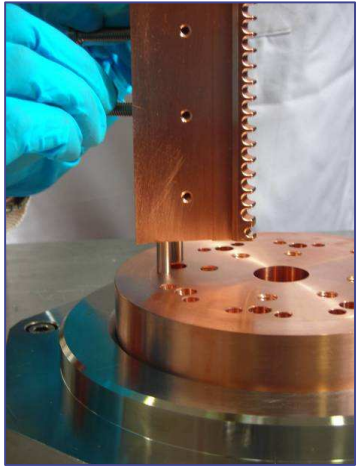

Figure 4. Positioning of the copper rods during assembly

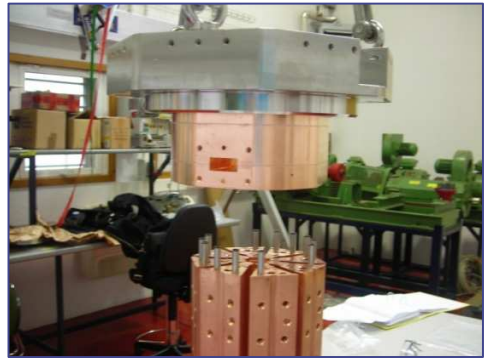

Figure 5. Positioning of the coupler during assembly.

*2.1.3. Low power RF tests.*
To test the whole structure, three different test benches were used for the first prototype: single copper rod, power extractor and PETS.

a) Single bar test bench. A special test bench was designed to check the shape accuracy by RF

measurements on each rod. It consists on two side blocks put together with a single PETS bar to create inside a mode with same properties as the decelerating mode. Results agreed with 3-D mechanical measurements, but it was not used in the rest of the bars due to the risk of scratching on the rod side faces.

b) Power extractor test bench. The S-parameters were measured using two custom input couplers that produce the nominal $TM_{01}$-like mode within the PETS. Results were in agreement with the simulations. At 12 GHz, the measured parameters were: $S_{11}$= - 40 dB (reflection), $S_{21}$= - 3 dB (transmission through one waveguide) and $S_{31}$= - 3 dB (transmission through the other waveguide).

c) PETS structure test bench. Firstly, S-parameters measurements were taken on the PETS structure, obtaining the following results: $S_{11}$= -16 dB (reflection), $S_{41}$= -36 dB (transmission to end port at the beam pipe) and $S_{21}=S_{31}$= -3.8 dB (extraction through waveguide).

Additionally, several measurements with an antenna were taken, measuring the electric field through the slots between the rods. In that, several phase shift measurements for different frequencies were taken, obtaining that the detuning was about 40 MHz. But this value was not the real one, since PETS must be synchronized with a relativistic electron beam. To do that, it was necessary to compare the PETS dispersion curve with the speed of light line (Figure 6), obtaining a real detuning about 50 MHz.

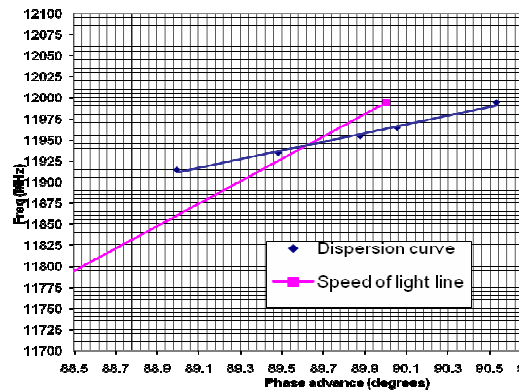

Figure 6. Dispersion curve vs speed of light line.

## 2.2 PETS TBL series production.

Before starting the series production, some improvements in the TBL PETS design were implemented. The most important modification was the introduction of ceramic damping plates, placed at both sides of the copper rods in order to absorb the HOMs generated by the interaction of the beam with the structure. In this case, the HOMs absorbers should have the following dielectric properties at PETS working frequency: dielectric constant about 20-30 and loss tangent at least 0.3. An exhaustive market survey was performed at CIEMAT [8], and after testing at CERN, the selected material was the SiC from ESK, reference EkaSiC-F [9].

Additionally, other modifications were implemented for improving the mechanical design and assembly:
- The dowel pins for copper rods alignment were moved to improve accuracy.

- A unique steel flange for waveguide was designed, used both for RF connection and closing the vacuum tank.
- The water cooling circuit and the connector design were modified to avoid that a potential leak directly pours into the vacuum of the beam.

The CIEMAT contribution to the series production was the assembly, low power testing of three new PETS, and the fabrication and brazing of several pieces such as cooling pipes, waveguides and flanges for seven PETS which were installed in Autumn 2011 and for four tanks more, which will be placed in the line at the beginning of 2012.

## 3   Double Length CLIC PETS.

The Double Length CLIC PETS were designed by CERN to generate RF power in the CLEX module [10] and it has a length 2 times longer than CLIC PETS. It is based on a previous 11.4 GHz PETS tested at SLAC, although this PETS will be operated at the CLIC nominal frequency (12 GHz). The main characteristic of this PETS is its compact design, that includes a vacuum mini tank and a compact coupler. The compact coupler has the functions of power extraction and combination into a single waveguide.

The CIEMAT contribution includes the engineering design, fabrication, assembly and low power RF tests. Nowadays CIEMAT is still working on the assembly, so the RF low power tests are not reported here.

### 3.1   Engineering design.

Figure 7 shows an artistic view of the Double Length CLIC PETS. Similarly to the TBL PETS, it consists of eight OFE Cu rods with SiC plates fixed by aluminum supports. The copper rods have 80 regular cells and are joined by electron beam welding (EBW).

At both ends of the rods, two compact couplers are placed to allow the power recirculation. The couplers have several components and functions:
-A double choke to reflect the power with bandwidth broader than that in single choke.
-Two output waveguides that are combined in one, with a T-junction.
-A cooling circuit with a cooling capacity about

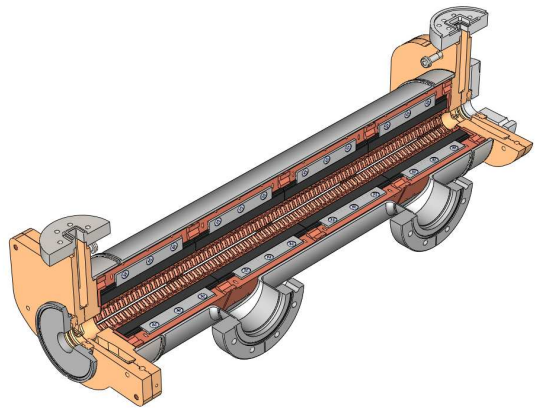

Figure 7. Artistic view of the double length CLIC PETS

LCWS11

100 W coming from the RF and beam losses.
- Flanges. They depend on the male or female compact coupler. The male coupler has the X-band RF male flange and a quick CF flange for the beam pipe, whereas the female coupler has the RF female flange, but the vacuum flange is of CF 63 type.

The joint between the couplers and bars is only by mechanical contact, supported by the outer vacuum tank, so the length of the tank is adjusted for guaranteeing the electric contact. The mini tank is joined by arc welding (TIG) to an intermediate stainless steel ring brazed to the coupler. The vacuum ports are made by pulling out the tank wall, since rounded edges are better to achieve ultra high vacuum operation.

### 3.2 Fabrication and assembly.

a) Copper rods. The rods are made of OFE Cu, 532 mm long, and similarly to the rods of the TBL PETS, the superficial and dimensional tolerances are very tight, i.e. roughness less than 0.4 μm and ±15 μm in shape. To achieve these tolerances, high speed milling using diamond tools with two intermediate stress relieve heat treatments were carried out. The bars (figure 8) are within tolerances, as figure 9 shows.

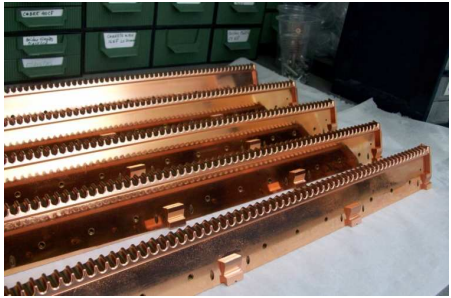
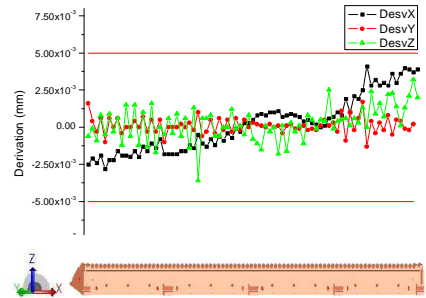

Figure 8. Double Length CLIC PETS copper bars.

Figure 9. Dimensional control of one of the bars of the Double Length CLIC PETS.

The bars also have SiC HOM absorbers of EkaSiC. In order to guarantee their complete degasification, the plates were fired at 1000ºC under vacuum below $10^{-5}$ mbar for one hour.

b) Compact coupler. The machining of the OFE Cu parts of the compact couplers was also very demanding. The RF surfaces had a roughness tolerance less than 0.3 μm in all pieces, whereas the dimensional tolerances were different depending on the part: ± 10 μm in the beam hole, ± 15 μm in the choke and ± 20 μm in the waveguide areas. So, the tools used for this step were high precision diamond monocrystal ones, and two intermediate stress relieve heat treatments were also necessary.

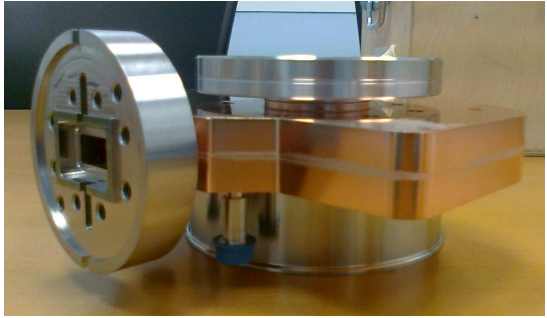

Figure 10: Brazed female compact coupler.

The joint of the compact couplers is one of the most critical steps in the fabrication, since it involves a lot of joints and dissimilar materials (Figure 10 shows the female compact coupler after brazing). The joint was performed by vacuum brazing in two steps. In the first step, the horizontal surfaces were brazed. After first brazing, the brazing surface of the coupler/RF flange joint was machined and then this joint was performed at lower temperature than that the first brazing step.

Table III summarizes the brazing parameters of the compact couplers.

Table III. Parameters for the brazing of the compact couplers.

| Brazing step | Type of joint | Filler alloy (%wt) | Filler melting range (ºC) | Vacuum (mbar) |
|---|---|---|---|---|
| First | Cu/Cu<br>Cu/SS (unplated) | 58Ag/32Cu/10Pd | 824-852 | <10-6 |
| Second | Cu/SS (unplated) | 68.5 Ag/26.5 Cu/5Pd | 807-810 | <10-6 |

c) Assembly. Assembly is the final step in the fabrication. It involves the joint of the copper rods and the final assembly of the structure by TIG welding of the vacuum mini-tank to the couplers. As it was exposed before, the weld joint preparation has been designed for guaranteeing the mechanical and electrical contact between the bars and the couplers. The weld preparation (figure 11) was designed as follows:
- Edge joint (square –lip) for reducing distortion in the rest of the structure.
- 0.05 Gap. It guarantees the bars/couplers mechanical contact during the pre-assembly before welding.
- 0.2 Gap. It guarantees that the maximum contraction during welding is 0.4 (adding both sides of the tank).

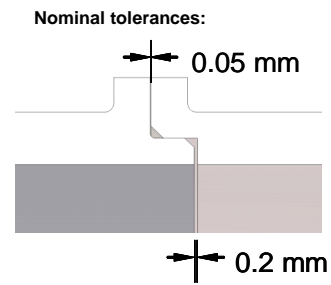

Figure 11: Weld preparation of tank/intermediate cap joint.

## 4   Conclusions.

CIEMAT has been collaborating with CERN on the PETS development since 2007.

First step was the engineering design, fabrication and RF tests of a PETS tank for the TBL experiment. The fabrication of the long copper rods (800 mm) with tight tolerances was specially challenging.

After successful installation of two prototypes at TBL line and deceleration measurements with 14 A and 19 A beams, the third and fourth PETS were sent to CERN and also installed in the line.

Finally, the fabrication of a double length CLIC PETS is ongoing, which will be part of the Test Module to demonstrate the two beam acceleration concept as proposed for CLIC. In this case, a new compact coupler design is implemented, with a mini-tank for vacuum tightness.

# 5  Acknowledgments.

This work has been partially funded by EU, VII Framework Programme, Grant Agreement 227579 and the Spanish Ministry of Science and Innovation under projects FPA2010-21456-C02-02 and FPA2007-29109-E.

Authors thank warmly the numerous people involved in the production: Utillajes Huerta, DMP and CIEMAT workshop for the machining of the parts; AIMEN, ITP, ECOR Research, and Serge Mathot (CERN) for the brazing, and Trinos Vacuum Projects for the vacuum tanks.

# 6  References.